\documentclass[twocolumn]{aastex631}
\usepackage{savesym}
\savesymbol{tablenum}
\usepackage{siunitx}
\restoresymbol{SIX}{tablenum}
\usepackage{amsmath}
\usepackage{multirow}
\usepackage[hang,flushmargin]{footmisc}

\graphicspath{{./}{figures/}}

\begin{document}
	\defcitealias{spilker20a}{S20}
	\defcitealias{spilker18}{S18}
	
	\title{On the Evidence for Molecular Outflows in High-redshift Dusty Star-forming Galaxies}
	
	\correspondingauthor{James Nianias}
	\email{nianias@hku.hk}
	
	\author[0000-0001-6985-2939]{James Nianias}
	\affiliation{Department of Physics, University of Hong Kong, Pokfulam Road, Hong Kong}
	
	\author[0000-0003-4220-2404]{Jeremy Lim}
	\affiliation{Department of Physics, University of Hong Kong, Pokfulam Road, Hong Kong}
	
	\author[0000-0002-5697-0001]{Michael Yeung}
	\affiliation{Max-Planck-Institut für extraterrestrische Physik, Giessenbachstraße, 85748 Garching, Germany}
	
	
	\begin{abstract}
		
		Galactic-scale outflows of molecular gas from star-forming galaxies constitute the most direct evidence for regulation of star formation.  In the early universe ($ z > 4 $), such outflows have recently been inferred from gravitationally-lensed dusty star-forming galaxies (DSFGs) based on ubiquitous detections of OH absorption extending to more blueshifted velocities than [CII] or CO emission in spatially-integrated spectra.  Because these lines are redshifted to sub-mm wavelengths, such measurements require careful corrections for atmospheric absorption lines, and a proper accounting of sometimes large variations in measurement uncertainties over these lines.  Taking these factors into consideration, we re-analyze OH and [CII] data taken with ALMA for the five sources where such data is available, of which four were categorised as exhibiting outflows.  Based on their spatially-integrated spectra alone, we find statistically significant ($ \geq 3 \sigma $) OH absorption more blueshifted than [CII] emission in only one source.  By contrast, searching channel maps for signals diluted below the detection threshold in spatially-integrated spectra, we find evidence for a separate kinematic component in OH absorption in all five sources in the form of: (i) more blueshifted OH absorption than [CII] emission and/or (ii) a component in OH absorption exhibiting a different spatio-kinematic pattern than [CII] emission, the latter presumably tracing gas in a rotating disc.  Providing a more complete and accurate assessment of molecular outflows in gravitationally-lensed DSFGs, we suggest methods to better assess the precision of corrections for atmospheric absorption and to more accurately measure the source continuum in future observations.
		
	\end{abstract}
	
	\section{INTRODUCTION}
	\label{section:intro}

	\begin{figure*}[t]
		\includegraphics[width=7in]{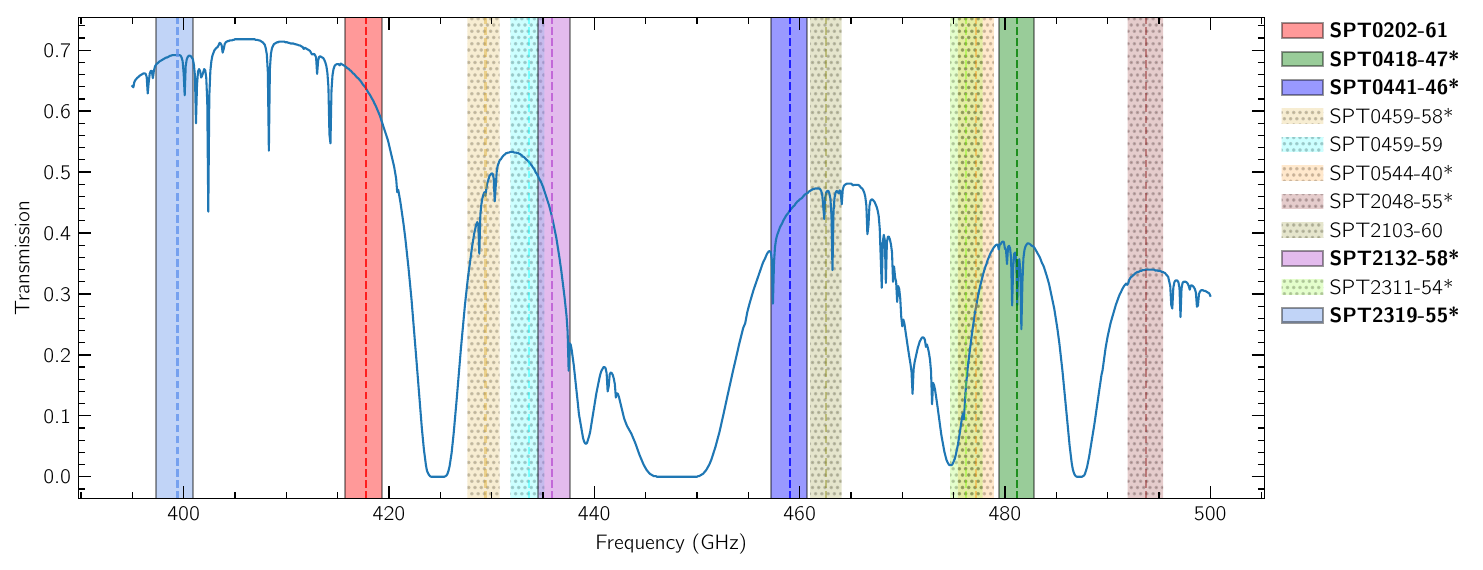}
		\caption{OH-containing bandpasses of the sources observed in \citetalias{spilker18} and \citetalias{spilker20a} (coloured bands) overlaid on a representative atmospheric transmission model (blue curve) produced in CASA \citep{casaref}.  Sources with high-resolution [CII] observations are indicated in bold, while those in which \citetalias{spilker18} or \citetalias{spilker20a} detect blueshifted OH absorption are indicated with asterisks.  Dashed vertical lines indicate the frequency of the higher-frequency OH line at the respective source redshift.  The broad absorption bands in the transmission model are produced by H$ _2 $O, while the narrow lines are due to O$ _3 $.  The bandpasses of SPT0418-47 (green band) and SPT2319-55 (light blue band) contain O$ _3 $ atmospheric absorption lines, while the atmospheric transmission drops steeply towards the high-frequency (low-velocity) edge of the bandpass of SPT2132-58 (purple band).}
		\label{fig:bandpasses}
	\end{figure*}
	
	Feedback processes related to star formation, active galactic nuclei, or both, are thought to have regulated the formation of stars in galaxies since early times.  Without feedback, cosmological models predict galaxies to have much larger stellar masses than are seen in the present-day universe (e.g., see \citealt{keres09} and references therein).  The most easily visible manifestation of such feedback is galactic-scale outflows, a prominent example of which can be seen in the nearby starburst galaxy M82.  This outflow is detectable in X-ray, H$ \alpha $, and CO, tracing ionised, atomic, and molecular gas, along with infrared emission tracing dust, all streaming outwards perpendicular to the disc.  Since young galaxies can exhibit more vigorous star formation, outflows at high redshift may be even stronger \citep{sugahara19}; furthermore, galaxies have lower masses in the early universe, so both gas and dust can escape more easily from the galaxy owing to radiation pressure and stellar winds from star-forming regions.
	
	Unlike in nearby galaxies, galactic outflows at high redshifts are generally not spatially resolved and are instead inferred from emission and/or absorption line profiles.  High-redshift star-forming galaxies have been shown to exhibit spectral signatures of ubiquitous ionized/atomic outflows: Ly$ \alpha $ emission lines with peaks redshifted relative to the systemic velocity of the host galaxy, sometimes accompanied by weaker blueshifted peaks, are interpreted in terms of scattering of Ly$ \alpha $ through partially- or fully-ionized outflows (see reviews by \citealt{ouchi20}, \citealt{erb15}), though radiative transfer models are as yet unable to reproduce the observed  of different Ly$ \alpha $ profiles \citep{gl22}.  A more unambiguous indicator of outflows is the presence of blueshifted metal absorption lines arising from partially neutral material outflowing from starburst regions towards the observer \citep[see][]{steidel10}, though such features are difficult to detect when the stellar continuum is weak.
	
	The apparent abundance of ionized/atomic outflows in star-forming galaxies at high redshifts naturally raises the question of whether molecular gas also is driven out.  Molecular gas is of particular interest because it comprises the fuel for star formation, making outflows of molecular gas the most direct signature for feedback-regulated star formation in star-forming galaxies.  At the present time, however, high-redshift molecular outflows seen in line emission, as inferred from broad CO line wings, have only been reported in a small number of objects -- around ten -- all thought to be AGN-driven (e.g., review by \citealt{veilleux20}).  Given the limited success in observing high-redshift molecular outflows in line emission, observers have resorted to searching for outflows in line absorption, since all that is required is a sufficiently bright background continuum source such as dust emission.  \cite{spilker18} and \cite{spilker20a} (hereafter \citetalias{spilker18} and \citetalias{spilker20a}, respectively) took advantage of a sample of gravitationally-lensed galaxies with particularly bright dust continua -- dusty star-forming galaxies or DSFGs -- to search for spectral signatures of outflowing molecular gas traced in OH $ \SI{119}{\um} $ doublet absorption.  They observed eleven of these galaxies with the Atacama Large Millimeter/submillimeter Array (ALMA) in Band 8 (covering $ 385 $ -- $ \SI{500}{GHz} $, corresponding to a redshift range of $ z \sim 4 $ -- 5.7 in the OH $ \SI{119}{\um} $ doublet), and compared the resulting OH absorption spectra with archival observations of either [CII] or CO emission from the same objects.  In eight of the eleven sources observed, they detect OH absorption blueshifted to velocities greater than any detectable [CII] or CO emission.  Interpreting the [CII] and CO emission as tracing gas contained in the disc of the target galaxies, they argue that the blueshifted OH absorption traces outflowing gas.  Moreover, they find no strong evidence of AGN-related emission based on mid-IR photometry, suggesting that these outflows may be driven by feedback from star-formation.
	
	As was pointed out in both \citetalias{spilker18} and \citetalias{spilker20a}, observations in the sub-mm are complicated by both broad and narrow atmospheric absorption features.  These features can be seen in a representative atmospheric transmission curve for ALMA band 8, the band in which all the OH observations in \citetalias{spilker18} and \citetalias{spilker20a} were made, as plotted in Figure \ref{fig:bandpasses}.  The coloured bands in this figure indicate the bandpasses of each target from \citetalias{spilker18} and \citetalias{spilker20a}.  As can be seen, some of these bandpasses coincide with large changes in atmospheric transmission, either in the form of narrow lines due to O$ _3 $ (e.g., SPT2319-55 and SPT0418-47) or broad features across the bandpass due to H$ _2 $O absorption (e.g., SPT2132-58).  The strong and sometimes rapid changes in atmospheric absorption with observing frequency, as well as variations in time owing to changes in atmospheric conditions and/or source elevation, have two important consequences that obervers must consider.  Firstly, although corrections are made to such changes through measurements of atmospheric opacity towards each object observed (see Section \ref{section:calib}), it is still necessary to assess the precision of these corrections to ensure that no residual atmospheric features remain.  Secondly, varying atmospheric transmission with frequency leads to a varying noise level across the target spectra.  While this effect is mentioned in \citetalias{spilker18}, it is crucial to explicitly compute measurement uncertainties in each channel separately so as to assess the statistical significance of any spectral features.
	
	Here, we re-analyse all sources in \citetalias{spilker18} and \citetalias{spilker20a} for which high-resolution [CII] $ \SI{158}{\um} $ observations have been made with ALMA -- five out of eleven in total, as listed in Table \ref{table:observations}.  The remaining sources (all from \citetalias{spilker20a}) only have available observations in CO at angular resolutions far too low to permit the checks described above, and are therefore excluded from our study.  To address the concerns raised above, we check the precision of the overall bandpass calibration to define the minimum level whereby OH absorption features can be considered genuine.  In so doing, we find that in all but one case, the bandpass is corrected to a sufficiently high precision that any residual atmospheric effects must be well below the noise level of the target spectra.  We also determine the measurement uncertainty in each spectral channel, necessary to assess the statistical significance of any putative OH absorption.  As a first step in searching for outflows, we compare spatially-integrated OH and [CII] spectra in exactly the same manner as \citetalias{spilker18} and \citetalias{spilker20a} to search for OH absorption blueshifted beyond detectable [CII] emission.  Of the five sources included in our study, \citetalias{spilker18} and \citetalias{spilker20a} find such evidence for outflows in four; by contrast, with a proper accounting of measurement uncertainties in different spectral channels, our analysis reveals just one source in which this spectral signature is statistically significant.
	
	Owing to gravitational lensing, all of the sources studied are strongly magnified into Einstein rings, allowing other diagnostics to be used for assessing outflows.  Thus, secondly, to guard against the possibility of localised OH absorption and/or [CII] emission being diluted below the detection threshold in the spatially-integrated spectra, we examine maps in individual spectral channels (channel maps) for each source.  We find [CII] emission extending to more blueshifted velocities than is detectable in the spatially-integrated spectra in two sources, highlighting a major potential pitfall in relying on spatially-integrated spectra alone to diagnose outflows.  On the other hand, we find OH absorption extending to more blueshifted velocities than [CII] emission not just in the source mentioned above showing such behavior in its spatially-integrated spectrum, but also in two further sources where the corresponding features are not detectable in their spatially-integrated spectra.  Finally, we search the channel maps for a component in OH absorption exhibiting a different spatio-kinematic pattern than the [CII] emission, thus providing direct evidence for a separate kinematic component possibly tracing an outflow, finding three such examples.
	
	In Section \ref{section:obs}, we describe our reduction of the data taken by ALMA, focusing in particular on how well the effects of atmospheric absorption are accounted for in the calibration process.  We stress that we follow the same steps in the data reduction as \citetalias{spilker18} and \citetalias{spilker20a}, and obtain the same results within random noise fluctuations -- except now explicitly quantifying the accuracy to which corrections in atmospheric absorption have been made in the target sources as well as quantifying the level of noise fluctuations in each spectral channel.  In Section \ref{section:results}, we compare the spatially-integrated [CII] and OH spectra of each source as well as the spatio-kinematic structure of [CII] and OH in maps of individual spectral channels.  In Section \ref{section:discussion}, we summarise the complete suite of evidence for outflows in each source.  We then discuss ramifications for the mass outflow rates as calculated by \cite{spilker20b}.  We end our discussion with suggestions to mitigate the difficulties in detecting weak astronomical absorption lines in the presence of strong atmospheric absorption lines in ALMA observations.  Finally, in Section \ref{section:sum}, we summarise our work and offer our thoughts on how future searches of OH outflows from high-$ z $ galaxies ought to be conducted.
	
	\section{ARCHIVAL OBSERVATIONS AND DATA REDUCTION}
	\label{section:obs}
	
	We obtained data for the same OH $ \SI{119}{\um} $ observations presented in \citetalias{spilker18} and \citetalias{spilker20a}, along with data from ancillary observations covering the [CII] $ \SI{158}{\um} $ line of each target, from the ALMA archive.  The OH observations were made with one sideband containing two partially-overlapping spectral windows covering the OH lines at the source redshift, and an alternate sideband containing two partially-overlapping spectral windows covering only the dust continuum.  Further details of the OH observations can be found in \citetalias{spilker18} and \citetalias{spilker20a}. Similarly, the [CII] observations utilised two spectral windows to cover the [CII] line at the source redshift in one sideband, and two spectral windows to cover the dust continuum in the other sideband.  In both the OH and [CII] observations, each spectral window is configured at its maximal possible bandwidth of $ \SI{1.875}{GHz} $.  Table \ref{table:observations} lists the redshifts and integration times of the observations in OH and [CII] for each source.
	
	\begin{table}
		\caption{Sources}
		\begin{tabular}{c c c c}
			\hline \hline
			Source & $ z $ & \multicolumn{2}{c}{int. time (min)} \\ 
			&  & OH & [CII] \\ [0.5ex]
			\hline
			SPT0202-61 & 5.0180 & 43 & 54 \\ 
			
			SPT0418-47 & 4.2248 & 87 & 19 \\
			
			SPT0441-46 & 4.4770 & 28 & 29 \\
			
			SPT2132-58 & 4.7677 & 17 & 25 \\
			
			SPT2319-55 & 5.2943 & 30 & 21 \\ [1ex] 
		\end{tabular}
		\label{table:observations}
	\end{table}
	
	\subsection{Calibration}
	\label{section:calib}
	
	\begin{figure*}
		\includegraphics[width=7in]{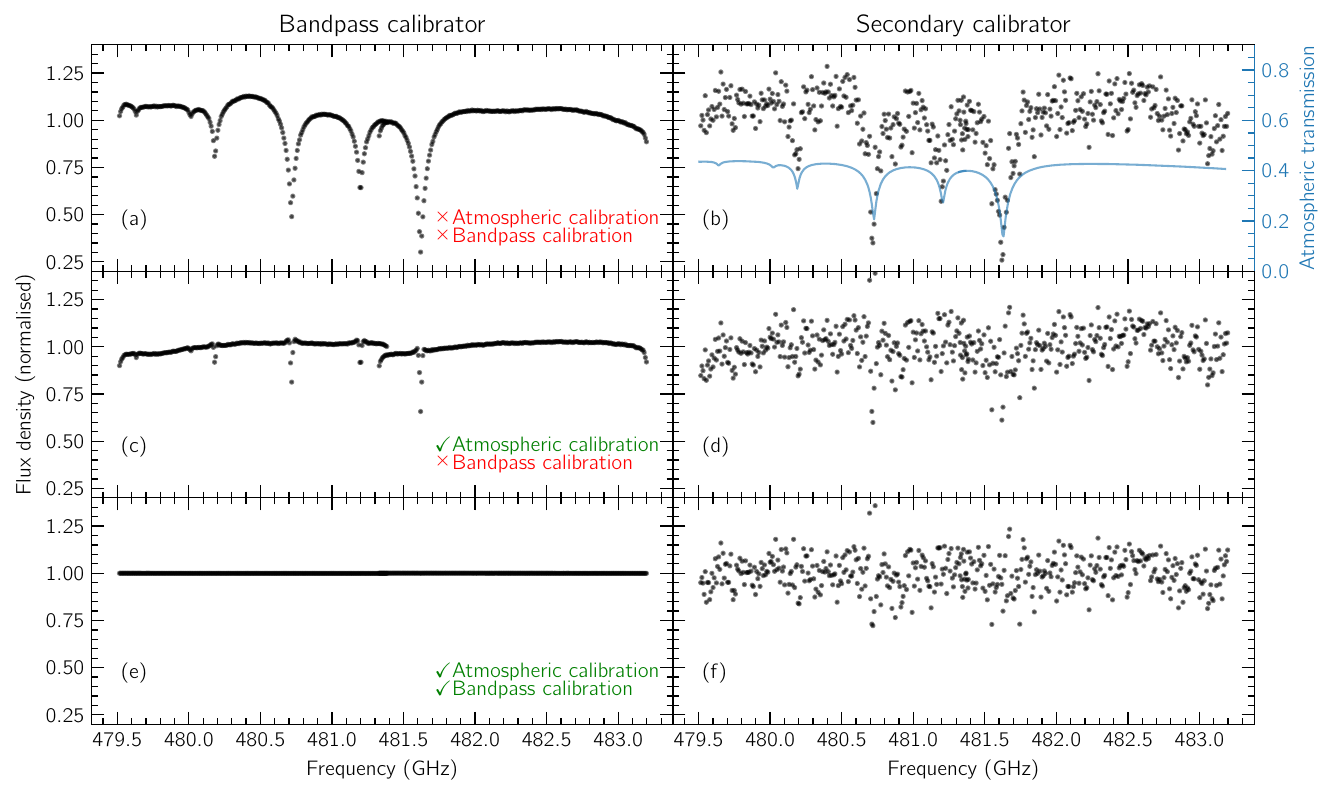}
		\caption{\textbf{(a)-(b):} Spectrum of the bandpass calibrator and secondary (phase) calibrator for SPT0418-47 with neither atmospheric nor instrumental bandpass calibration applied.  Each spectrum clearly shows narrow absorption lines due to atmospheric O$ _3 $, closely matching the atmospheric transmission model in (b) (blue curve). \textbf{(c)-(d):} the same spectra after atmospheric calibration but before instrumental bandpass calibration.  Both spectra cleary show residual features around the O$ _3 $ lines due to the lower spectral resolution of the atmospheric opacity measurements. \textbf{(e)-(f):} the same spectra after both atmospheric and instrumental bandpass calibration. Since the instrumental bandpass correction is derived from this spectrum, the bandpass calibrator spectrum in panel (e) is almost perfectly flat; however, the secondary calibrator spectrum in panel (f) shows enhanced scatter around the O$ _3 $ lines due to higher thermal noise, residual atmospheric features, or both.} 
		\label{fig:spt0418bandpass}
	\end{figure*}
	
	The ALMA Science Archive provides uncalibrated visibilities along with scripts that allow users to calibrate the visibilities using the corresponding pipeline for the ALMA cycle in which the observations were made.  We calibrated each visibility set in the same manner as reported in \citetalias{spilker18} and \citetalias{spilker20a}, using the calibration script either executed by us (for SPT0418-47 and SPT2132-58) or by the East Asia ALMA helpdesk (for the remaining sources\footnote{Although we ran the calibration pipelines ourselves for all the sources, owing to a number of technical issues, we had to revert to the ALMA helpdesk for calibrating the data for some of these sources.}).  The only exceptions are the OH data for SPT0441-46 and [CII] data for SPT2132-58, which we re-calibrated manually, though excluding the data flagged in the initial automatic pipeline as for the other sources\footnote{Our re-calibration of the OH observations of SPT0441-46 was motivated by concerns about the high noise level found in the images made with the original calibration; our re-calibrated data shows reduction in sidelobe levels.  For SPT2132-58, an issue in the original ALMA calibration script manifested as a discontinuous jump in noise level between two adjacent spectral windows.}.  Following subsequent self-calibration of the visibilities (described in Section \ref{section:selfcalclean}), the dust continuum maps that we produced for all sources have morphologies and signal-to-noise ratios in close agreement with those reported in \citetalias{spilker18} and \citetalias{spilker20a}; furthermore, the spatially-integrated OH spectra also are similar to those presented in \citetalias{spilker18} and \citetalias{spilker20a} within random noise fluctuations.
	
	Prior to calibration, the atmospheric absorption features shown in Figure \ref{fig:bandpasses} are imprinted on the spectra of the target as well as the calibrators.  These features are clearly visible in the uncalibrated spectra of the bandpass and secondary calibrators (the purposes of these calibrators are described below), examples of which are shown in Figure \ref{fig:spt0418bandpass}(a) for the pimary calibrator and Figure \ref{fig:spt0418bandpass}(b) for the secondary calibrator used in the observations of SPT0418-47.  ALMA corrects for variable atmospheric transmission by measuring the signal strength emitted by the atmosphere, expressed as the atmospheric temperature $ T_{atm} $, which is then used to estimate atmospheric opacity and hence also transmission (for details see \citealt{he22}).  By making this measurement as a function of both frequency (at a resolution of $ \SI{16}{MHz} $) and time (immediately before an observation of a calibrator or the target object), the calibration pipeline corrects for changes in the atmospheric transmission due to the changing elevation of the target as well as varying weather conditions.  Because measurements of $ T_{atm} $ are obtained at lower spectral resolution than the calibrators and target spectra, however, residual atmospheric features remain even after the atmospheric calibration is applied.  Such residuals are most readily apparent for relatively narrow atmospheric absorption lines that are not spectrally resolved in the $ T_{atm} $ measurements, as can be seen in Figure \ref{fig:spt0418bandpass}(c) for the bandpass calibrator and Figure \ref{fig:spt0418bandpass}(d) for the secondary calibrator after atmospheric calibration is applied.
	
	The remaining gradual variation with frequency bracketed by roll-offs towards the edges of each spectral window, most clearly apparent in Figure \ref{fig:spt0418bandpass}(c) and also present in Figure \ref{fig:spt0418bandpass}(d), is due to the instrumental response.  To correct for this, an observation of a bright continuum source (referred to as the bandpass calibrator, which has no known spectral lines) is conducted to derive corrections for the instrumental spectral response; as there are residual atmospheric features in the spectrum of the bandpass calibrator, these features are also removed in the bandpass calibration.  Figure \ref{fig:spt0418bandpass}(e) shows the spectrum of the bandpass calibrator after bandpass calibration is applied: as expected, it is perfectly flat to within the noise level.  On the other hand, the bandpass calibration does not necessarily remove (and may even exacerbate) the residual atmospheric features in the spectrum of the secondary calibrator and target, which are often observed at different (lower) elevations than the bandpass calibrator and therefore suffer different (larger) degrees of atmospheric absorption.  Figure \ref{fig:spt0418bandpass}(f) shows the spectrum of the secondary calibrator after bandpass calibration is applied, in which channels coinciding with the narrow atmospheric O$ _3 $ lines can be seen to exhibit enhanced scatter owing to either residual atmospheric features or higher thermal noise in channels with higher atmospheric opacity (or a combination thereof).  This check highlights the importance of determining the precision of the combined atmospheric and bandpass calibration to ensure that any residual atmospheric features are sufficiently small that they cannot significantly affect the target spectra.
	
	The final step in the calibration is to derive temporal variations in amplitude and phase induced by both the atmosphere and the telescope instrumentation on the target object.  Such corrections are derived from each scan of the secondary calibrator, and interpolated to the times of the target scans so as to be applied to the target.
	
	Finally, the measurement uncertainty can change quickly across the spectrum of a given object owing to rapid changes in atmospheric transmission with frequency.  Such rapid changes in the measurement uncertainty need to be recognised when assessing the statistical significance of any purported spectral line features, especially when the latter coincide with narrow and deep atmospheric absorption lines (e.g., as in the cases of SPT0418-47 and SPT2319-55 as described in Section \ref{section:intspecs}).
	
	\begin{figure*}
		\includegraphics[width=7.5in]{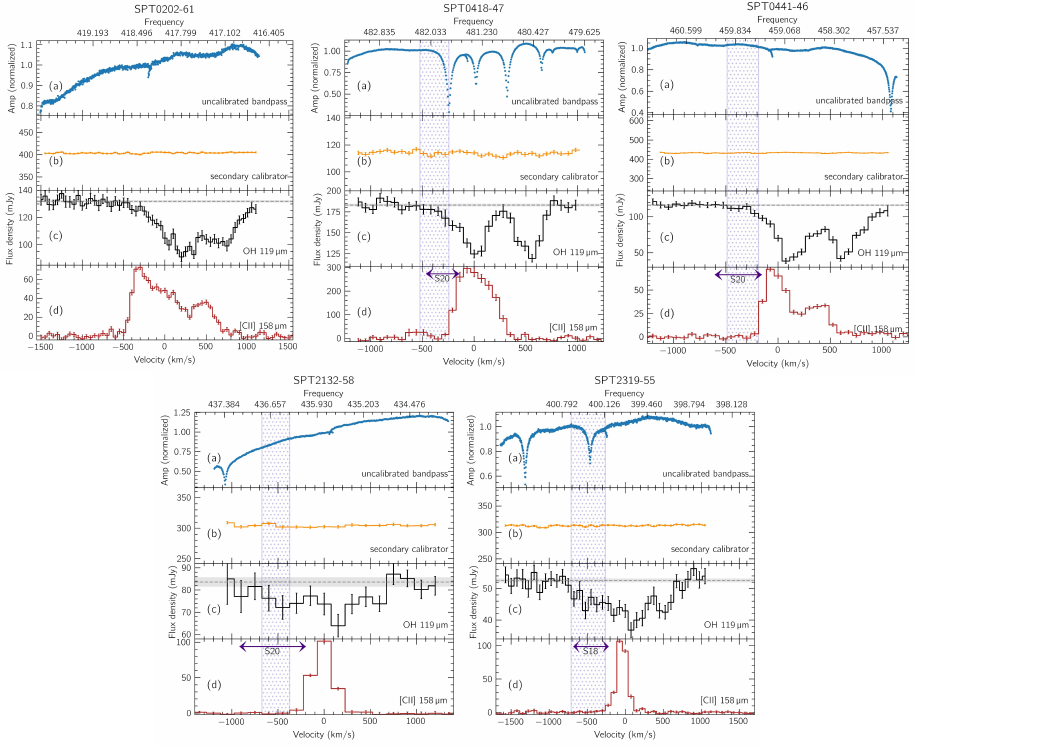}
		\caption{\textbf{(a)} Bandpass calibrator spectra without atmospheric or bandpass corrections, showing the dominant imprints of changing atmospheric transmission along with instrumental bandpass.  \textbf{(b)} spectra of the secondary calibrators after calibration, showing no statistically significant residual features.  \textbf{(c)} spatially-integrated OH spectra from each source, with grey horizontal bars indicating the $ \pm 1\sigma $ level of the continuum, blue hatched regions indicating the velocity ranges over which we obtain the highest signal-to-noise OH absorption blueshifted beyond the [CII] emission, and double-headed purple arrows indicating the approximate ranges over which \citetalias{spilker18} or \citetalias{spilker20a} report blueshifted OH absorption.  The $ \pm 1 \sigma $ error bars of the OH spectra are shown for each channel, and can be seen to increase in size with decreasing atmospheric transmission.  \textbf{(d)} [CII] spectra extracted over the same regions as OH, and binned to the same velocity resolution.}
		\label{fig:spectra}
	\end{figure*}
	
	\subsection{Self-calibration}
	\label{section:selfcalclean}
	
	Owing to their different sky positions and changing weather conditions, the amplitude and phase corrections as interpolated from the secondary calibrator may leave residual amplitude and phase errors in the calibrated data for the target.  Such residual errors can be reduced using self-calibration, which in the cases here utilises a high-signal-to-noise continuum map of the target itself to determine any remaining amplitude and phase corrections.  In nearly all of the observations, the spectral windows containing the OH or [CII] lines have wide enough frequency coverage that they also permit measurements of the continuum adjacent to the respective line profiles.  We therefore selected frequency ranges free from any OH absorption or [CII] emission, using the spectra shown in \citetalias{spilker18} and \citetalias{spilker20a} as a guide, and combined these with the continuum-only sidebands when making continuum images.  The sole exception is the OH observations of SPT2132-58, for which \citetalias{spilker20a} infer OH absorption to cover nearly the entire bandpass, and which has no useable data in the continuum sidebands due to low atmospheric transmission.  In this case, following \citetalias{spilker20a}, we used all channels in both OH-containing spectral windows to make the continuum image; as we will show in Section \ref{section:results}, most of these channels do not in fact contain statistically significant OH absorption.  We then CLEANed\footnote{The images generated are convolved with the point spread function (PSF) of the antenna array, known as the dirty beam.  CLEANing deconvolves the dirty beam from the image and replaces it with a clean beam, a Gaussian ellipsoid fitted to the central lobe of the PSF.} the resulting continuum maps to three times the thermal noise level\footnote{We determined the thermal noise level by making maps with large shifts in phase center relative to the source positions.} -- not so deep as to pick up noise peaks or artefacts -- for self-calibration.  We then used these CLEAN continuum maps to perform phase-only self-calibration of the target visibilities, which resulted in smaller residual phase errors and hence significantly improved signal-to-noise in the final CLEAN continuum maps.  Like \citetalias{spilker18} and \citetalias{spilker20a}, we did not perform any self-calibration in amplitude, as deriving robust amplitude corrections would require images having much higher signal-to-noise ratios.
	
	\subsection{Channel maps and integrated spectra}
	\label{section:chan_spec}
	
	The channel maps used in our study as presented in Section \ref{section:chanmaps} were generated from the self-calibrated visibilites.  For each source, we CLEANed the channel maps to $ 1.5 $ times the thermal noise level as measured in the channel map corresponding to the central velocity of the higher-frequency (lower-velocity) component of the OH doublet.  As a check on the quality of our CLEAN maps, we compute the root-mean-square (rms) variation over an annulus surrouding the source in each corresponding channel map, finding it to be only $ \sim 10\% $ to $ 30\% $ higher than the thermal noise level.  Table \ref{table:results} lists the OH channel widths used for the individual channel maps of each source, chosen to be approximately the same as those in \citetalias{spilker18} and \citetalias{spilker20a}.  To make a proper comparison of the OH and [CII] spectral profiles, we used the same widths when making [CII] channel maps for each source.  To generate continuum-subtracted [CII] channel maps, like in \citetalias{spilker18} and \citetalias{spilker20a}, we performed continuum subtraction on the [CII] visibilities in the uv-plane, fitting the continuum level using the same channels in the [CII]-containing spectral windows as were used to make continuum maps, i.e., channels well away from any emission.  We performed continuum subtraction in the same manner on the OH visibilities to make continuum-subtracted OH maps; when presenting the spatially-integrated spectra in OH, however, we retain the continuum to allow the depth in OH absorption relative to the continuum to be computed.  By comparing the OH absorption depth to the precision in atmospheric and bandpass calibration for a particular source, we are able to assess whether this absorption genuinely arises in the source.  Finally, we convolved the OH channel maps, continuum-subtracted [CII] channel maps, and dust continuum maps of each target source to the smallest possible common angular resolution (synthesised beam), which is listed in Table \ref{table:results} for each object.  After convolution, the final beam sizes are only slightly larger than the beam sizes reported in \citetalias{spilker18} and \citetalias{spilker20a} for the OH observations.
	
	\begin{table*}
		\begin{center}
			\caption{Results from spatially-integrated spectra}
			\begin{tabular}{p{0.7in} p{0.7in} p{0.5in} p{0.5in} p{0.5in} p{0.9in} p{1.0in} p{0.7in}}
				\hline \hline
				Source & Channel width & BMAJ & BMIN & BPA & Outflow range & Absorption strength & Correction accuracy \\ 
				& (\SI{}{km.s^{-1}}) & ($ \arcsec $) & ($ \arcsec $) & ($ ^\circ $) & (\SI{}{km.s^{-1}}) & (\SI{}{mJy.km.s^{-1}}) & (\SI{}{mJy.km.s^{-1}}) \\ [0.5ex]
				\hline
				SPT0202-61 & 50 & 0.48 & 0.42 & 63 & N/A & N/A & N/A  \\ 
				
				SPT0418-47 & 70 & 0.55 & 0.35 & 89 & $ [-525, -245] $ & $ 2320 \pm 980 $ ($ 2.4 \sigma $) & 890 \\
				
				SPT0441-46 & 75 & 0.35 & 0.26 & 79 & $ [-462, -187] $ & $ 1690 \pm 620 $ ($ 2.7 \sigma $) & 200 \\
				
				SPT2132-58 & 150 & 0.43 & 0.36 & 114 & $ [-675, -375] $ & $ 2820 \pm 1170 $ ($ 2.2 \sigma $) & 380 \\
				
				SPT2319-55 & 75 & 0.40 & 0.32 & 64 & $ [-712, -262] $ & $ 2890 \pm 470 $ ($ 6.1 \sigma $) & 130 \\ [1ex] 
				\hline
			\end{tabular}
			\label{table:results}
		\end{center}
	\end{table*}
	
	As OH absorption can only arise from sightlines towards the dust continuum, we extracted spatially-integrated OH and [CII] spectra for each source by integrating over regions where continuum is detected above $ 3\sigma $ following the same procedure as used in \citetalias{spilker18} and \citetalias{spilker20a}.  To determine the $ 1\sigma $ uncertainties in individual channels of the integrated OH or [CII] spectra, we use the \textsc{Essence} software package \citep{tsukui23}.  \textsc{Essence} derives uncertainties in spatially-integrated flux density using the noise autocorrelation function, as measured from each channel map with the target source masked out.  Computing the uncertainties in this manner accounts for the spatial correlation in the channel maps induced by both the CLEAN (synthesised) beam as well as any residual amplitude and phase errors in the calibration\footnote{The noise calculated this way is in close agreement with a commonly-used approximation whereby the uncertainty in spatially-integrated flux in each channel is taken to be the rms noise level multiplied by the square root of the number of beams within the region of integration (to within 5 -- 15\%, varying from source to source and channel to channel).}.  The spatially-integrated OH and [CII] spectra along with the varying $ 1\sigma $ uncertainties in each channel are shown in panels (c) and (d), respectively, for each target source in Figure \ref{fig:spectra}.  As can be seen by comparing the OH spectra with those of the un-corrected bandpass calibrators in panel (a) for each source in Figure \ref{fig:spectra}, the noise in each channel of the OH spectra depends on the atmospheric transmission, such that the noise level is higher where the atmospheric transmission is lower -- and can therefore vary considerably across the OH line profile (as is the case also in [CII] for the same reasons).
	
	Quantifying the level of line absorption requires knowledge of the continuum level.  To estimate the continuum level and its uncertainty in the spatially-integrated OH spectrum of all the target sources apart from SPT2132-58, we used the same channels in the OH-containing spectral windows that were used to generate the clean continuum map, i.e., channels well away from any absorption.  For the spatially-integrated OH spectrum of SPT2132-58, where the continuum maps were made using the full OH-containing bandpass, we estimated the continuum level using the channels at the low-frequency (i.e. high-velocity) end of the bandpass.  These channels are absorption-free according to the continuum estimate provided in \citetalias{spilker20a}.  We show the continuum levels by the dashed grey lines and their $ \pm 1\sigma $ uncertainties by the horizontal grey bands in panel (c) for each source in Figure \ref{fig:spectra}.
	
	\section{RESULTS AND ANALYSIS}
	\label{section:results}
	
	\subsection{Integrated spectra}
	\label{section:intspecs}
	
	In both \citetalias{spilker18} and \citetalias{spilker20a}, an outflow is indicated by OH absorption extending to blueshifted velocities beyond any detectable [CII] emission in spatially-integrated spectra.  Indeed, a comparison between the OH and [CII] spectra for all but one of the sources (SPT0202-61) shown in Figure \ref{fig:spectra} reveals systematic offsets below the continuum level at velocities lower than the lowest velocity at which [CII] is detected; furthermore, the nominal absorption depth in these channels is the same as that in \citetalias{spilker18} and \citetalias{spilker20a} to within measurement uncertainties.  To compute the statistical significance in the individual cases, we integrate the OH spectrum of each source over velocities bluewards of the lowest detectable [CII] velocity until the signal-to-noise level of this putative absorption is maximised.  The velocity ranges thus selected are shown by hatched blue bands in panels (c) and (d) for each source in Figure \ref{fig:spectra} except SPT0202-61, for which no systematic offset below zero is appreciable over the velocity range bluewards of the lowest detectable [CII] velocity.  For comparison, we also show the approximate velocity ranges over which \citetalias{spilker18} and \citetalias{spilker20a} report spectral signatures of OH outflows with double-headed purple arrows.  The integrated intensity of the blueshifted OH absorption and its related measurement uncertainty, along with the velocity range over which we perform this integration, are given for each source in Table \ref{table:results}.
	
	With a now complete accounting of the measurement uncertainties across the velocity range spanned by putative blueshifted OH absorption, we find statistically significant ($ \geq 3 \sigma $) evidence for blueshifted absorption in the spatially-integrated spectrum of only one source, SPT2319-55, at $ 6.1 \sigma $.  The source that shows the next strongest putative blueshifted OH absortion is SPT0441-46, at a significance level of $ 2.7 \sigma $, followed by SPT0418-47 at $ 2.4 \sigma $ and SPT2132-58 at $ 2.2 \sigma $.
	
	Before accepting as genuine any statistically significant blueshifted OH absorption, it is crucial to check that the calibration corrects the spectra to a sufficiently high precision that any residual atmospheric or bandpass features must be smaller than the depth of the putative OH absorption.  To do this, we plot the spectra of the secondary calibrators as shown in panels (b) of Figure \ref{fig:spectra}.  These calibrators were observed at a similar elevation but have a higher signal-to-noise in the continuum compared to the target sources.  After atmospheric and bandpass calibration (described in Section \ref{section:calib}), none of the secondary calibrator spectra exhibit statistically significant residual features.  Applying the same velocity ranges given in Table \ref{table:results} for each target to their respective secondary calibrator spectra, we compute $ 3\sigma $ upper bounds for any residual bandpass features as a fraction of the continuum; we can then be confident that the bandpasses of the target OH spectra have been corrected to within the same fraction of the continuum.  The upper bounds thus derived are given in terms of integrated flux (in $ \SI{}{mJy.km.s^{-1}} $) in Table \ref{table:results}.  Comparing the strength of the putative blueshifted absorption in each source with the limits to the accuracy of atmospheric corrections, we find that in most cases the former is roughly an order of magnitude higher than the latter, indicating that imperfections in bandpass calibration cannot be responsible for the systematic offsets below the continuum in the spectra of most sources.  The sole exception is SPT0418-47, for which the upper limit for residual bandpass features is comparable to the $ 1\sigma $ uncertainty in the blueshifted absorption, while the absorption strength itself is just $ 2.4\sigma $:  in this case there is a prominent atmospheric absorption line that partially overlaps with the putative blueshifted absorption.  While the secondary calibrator spectrum does not show any statistically significant residual feature corresponding to this line, we cannot rule out the possibility that residual calibration errors contribute in part to the measured absorption.

	\subsection{Channel maps}
	\label{section:chanmaps}

	\begin{figure*}
		\includegraphics[width=6in]{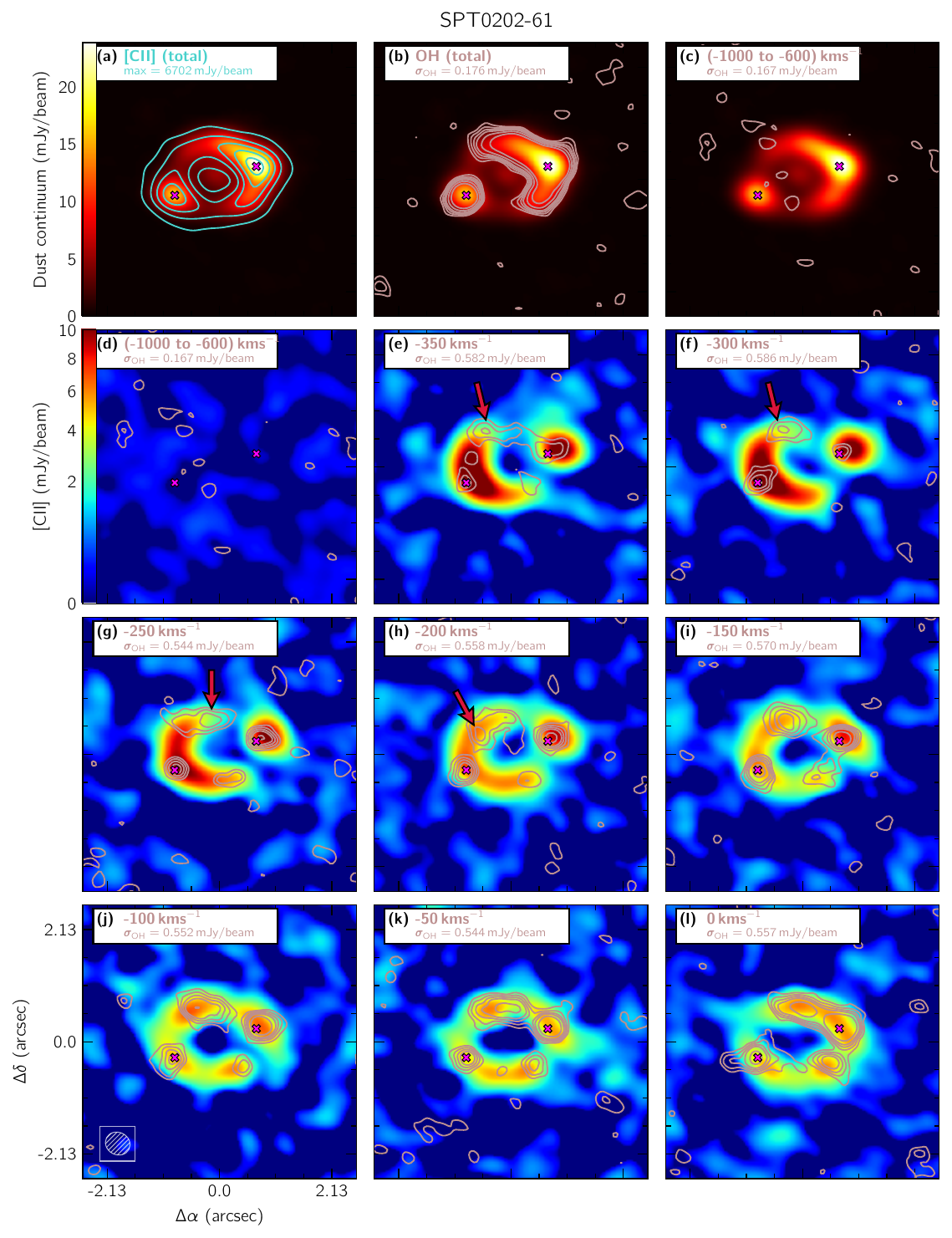}
		\caption{\textbf{(a)} Contours of total [CII] emission (moment 0) plotted at 10, 30, 50, 70, and 90\% of the peak surface brightness (given in the box in the upper left corner) against dust continuum from SPT0202-61 (seen as an Einstein ring). \textbf{(b)} contours of total OH absorption (moment 0) plotted at $ 2\sigma $, $ 3\sigma $, $ 4\sigma $, $ 5\sigma $, $ 7\sigma $, and $ 10\sigma $ against dust continuum ($ 1\sigma $ given in the box in the upper left corner, as in the remaining panels). \textbf{(c)} contours of OH absorption integrated over putative outflow channels (blue hatched range in Figure \ref{fig:spectra}) plotted at  $ 2\sigma $, $ 3\sigma $, $ 4\sigma $, $ 5\sigma $, and $ 7\sigma $ against dust continuum. \textbf{(d)} the same contours as (c), now overlaid on [CII] emission integrated over the same velocity range.  \textbf{(e) -- (l)} contours of individual OH velocity channel maps with widths of $ \SI{50}{km.s^{-1}} $ at central velocities as indicated in each panel plotted at  $ 2\sigma $, $ 3\sigma $, $ 4\sigma $, $ 5\sigma $, and $ 7\sigma $, and overlaid on [CII] emission channel maps over the corresponding velocity ranges (colour).  In all panels, magenta crosses mark the peaks of the dust continuum as shown in panels (a) to (c), and red arrows point to OH absorption features following a different spatio-kinematic trend to the [CII] emission, suggesting a separate kinematic component.}
		\label{fig:spt0202_velmaps}
	\end{figure*}
	
	\begin{figure*}
		\includegraphics[width=6in]{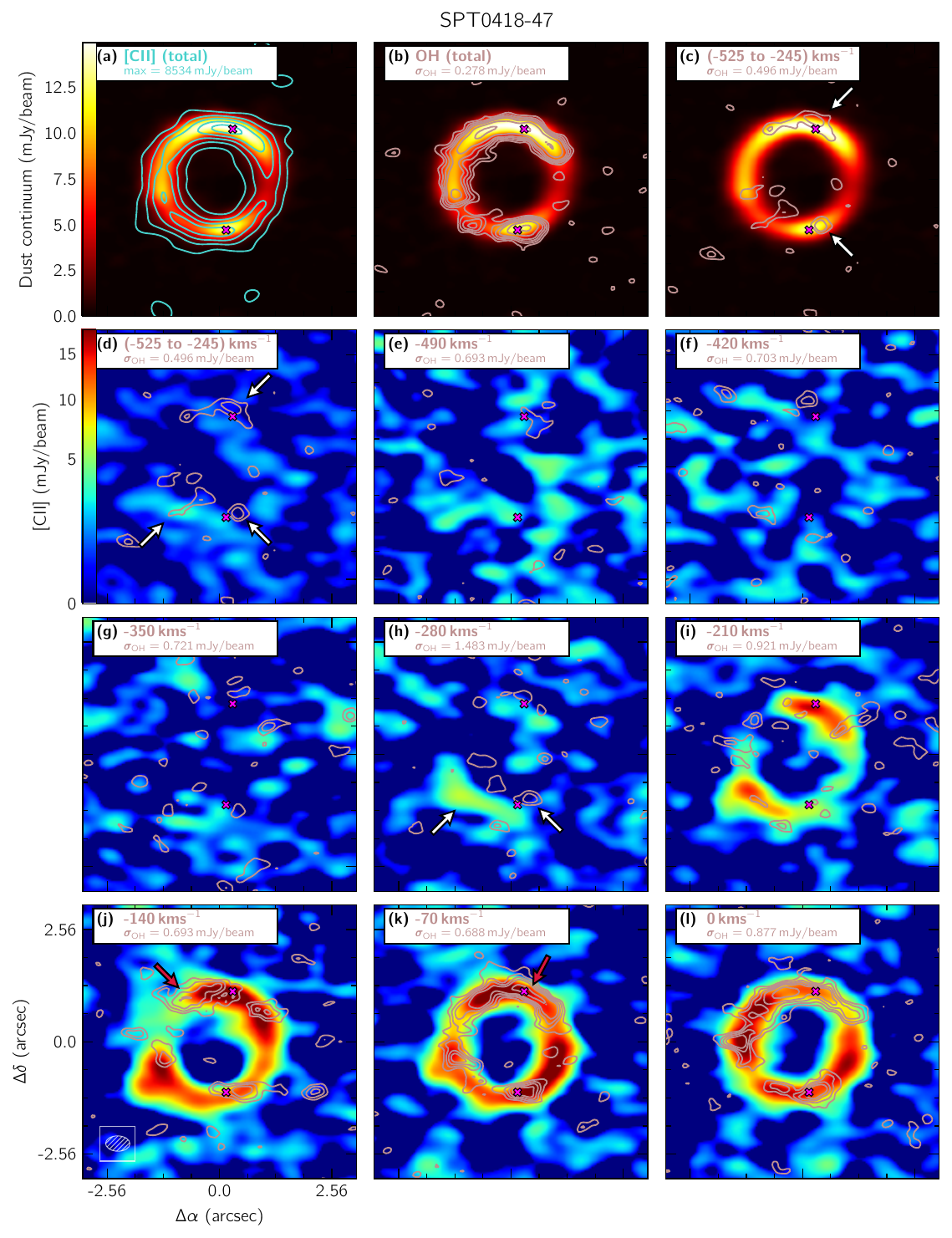}
		\caption{Same as figure \ref{fig:spt0202_velmaps}, but for SPT0418-47, with channel width of $ \SI{70}{km.s^{-1}} $.  White arrows in panels (d) and (h) point to patches of OH absorption or [CII] emission diluted to below the detection threshold in the spatially-integrated spectrum, while red arrows in panels (l) and (j) point to OH absorption features following a different spatio-kinematic trend to the [CII] emission, suggesting a separate kinematic component.}
		\label{fig:spt0418_velmaps}
	\end{figure*}
	
	\begin{figure*}
		\includegraphics[width=6in]{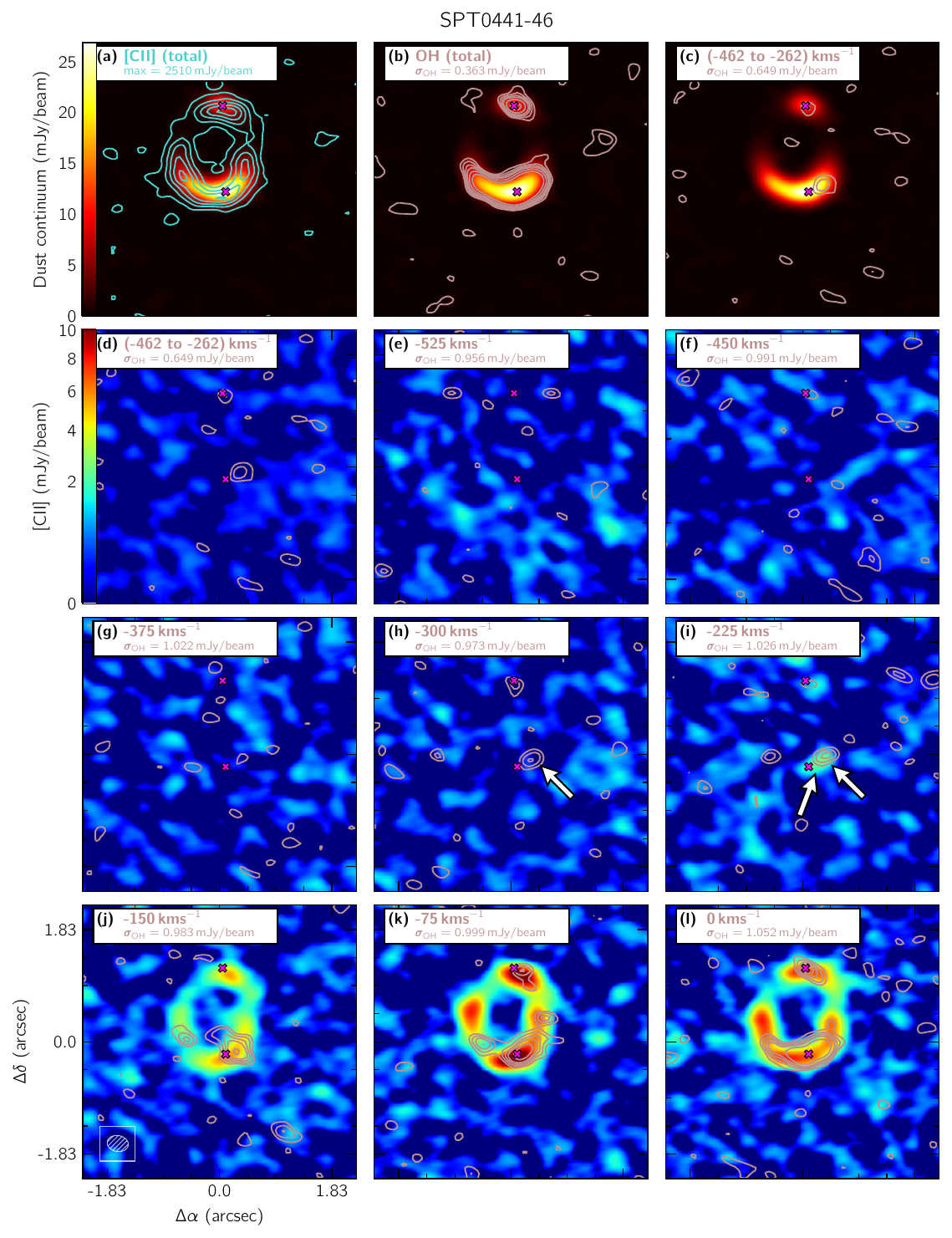}
		\caption{Same as Figure \ref{fig:spt0202_velmaps}, but for SPT0441-46, with channel width of $ \SI{75}{km.s^{-1}} $.  White arrows in panels (h) and (i) point to patches of OH absorption or [CII] emission diluted to below the detection threshold in the spatially-integrated spectrum, with panel (h) showing a patch of OH absorption blueshifted beyond any detectable [CII] emission.}
		\label{fig:spt0441_velmaps}
	\end{figure*}
	
	\begin{figure*}
		\includegraphics[width=6in]{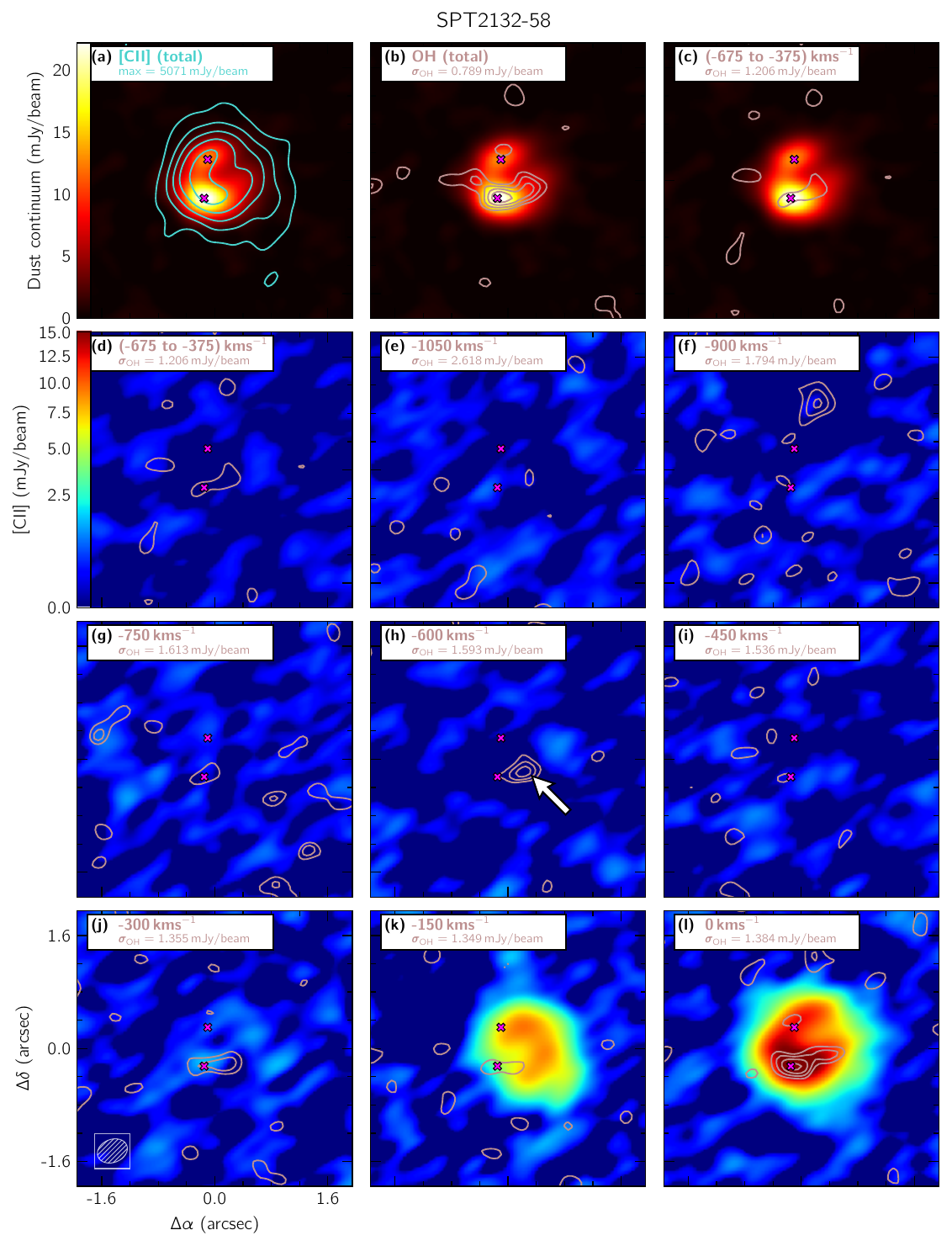}
		\caption{Same as Figure \ref{fig:spt0202_velmaps}, but for SPT2132-58, with channel width of $ \SI{150}{km.s^{-1}} $.  The white arrow in panel (h) points to a patch of OH absorption blueshifted beyond any detectable [CII] emission, which is diluted to below the detection threshold in the spatially-integrated spectrum.}
		\label{fig:spt2132_velmaps}
	\end{figure*}
	
	\begin{figure*}
		\includegraphics[width=6in]{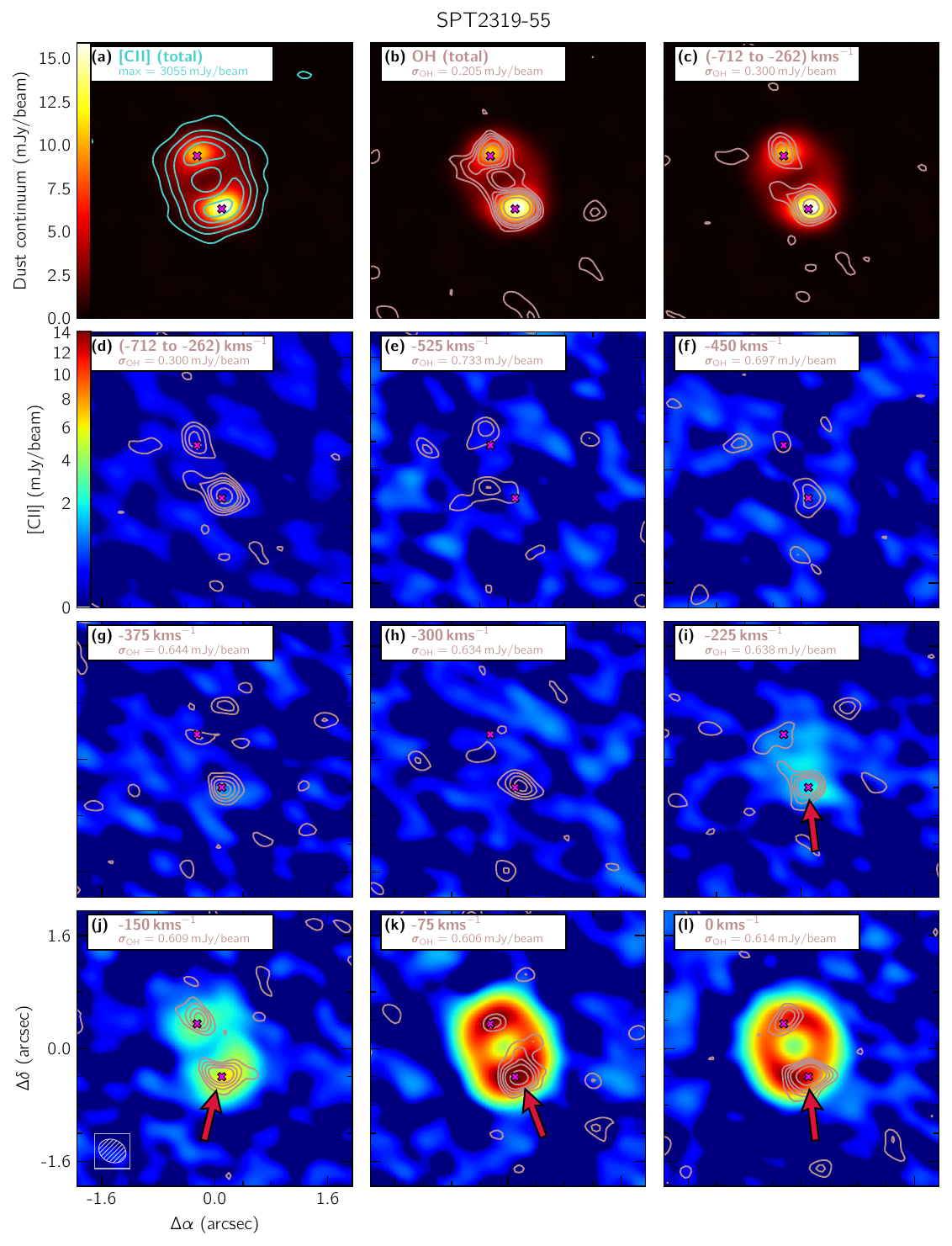}
		\caption{Same as Figure \ref{fig:spt0202_velmaps}, but for SPT2319-55, with channel width of $ \SI{75}{km.s^{-1}} $.  Patches of blueshifted OH absorption with no accompanying [CII] emission are clearly visible in channels (h) to (e).  In panels (l) to (j), red arrows indicate OH absorption features following a different spatio-kinematic trend to the [CII] emission, suggesting a separate kinematic component.}
		\label{fig:spt2319_velmaps}
	\end{figure*}
	
	Relying solely on the spatially-integrated spectra shown in Figure \ref{fig:spectra} to diagnose outflows raises two major potential pitfalls.  First, the process of spatial integration may dilute signals that are confined to localised regions of the Einstein ring, whether they be small patches of OH absorption or of [CII] emission.  Second, the possibility that [CII] emission extends to velocities beyond those detectable as dictated by the noise level can never be completely ruled out, making it ambiguous whether the detection of OH absorption more blueshifted than [CII] emission, alone, implies a kinematic component separate from the disc.  To address both these issues, we made both OH and [CII] channel maps of each source at the same velocity resolution as the spectra shown in Figure \ref{fig:spectra}, as well maps integrated over the full velocity ranges that we use to search for blueshifted OH (hatched blue bands in Fig. \ref{fig:spectra}), using the continuum-subtracted OH and [CII] visibilities.  These maps are shown in panels (b) to (l) of Figures \ref{fig:spt0202_velmaps} to \ref{fig:spt2319_velmaps} for each source in the same order as shown in Figure \ref{fig:spectra}, where contours correspond to OH absorption, red background to the dust continuum, and rainbow background to [CII] emission -- showing gravitationally-lensed images of each galaxy in the form of an Einstein ring.
	
	When analysing the channel maps, it is important to take into consideration apparently statistically significant ($ \ge 4\sigma $) patches of OH ``absorption" that are not spatially coincident with the dust continuum, as can be seen in Figures \ref{fig:spt0418_velmaps}(j), \ref{fig:spt0441_velmaps}(j), and \ref{fig:spt2132_velmaps}(f).  Such features may arise due to residual amplitude and phase errors left following calibration.  We therefore only consider as genuine those absorption patches that are not only spatially-coincident with the dust continuum, but that are also deeper than any apparent absorption patches that are not spatially-coincident with the dust continuum.
	
	White arrows in the channel maps of Figures \ref{fig:spt0202_velmaps} to \ref{fig:spt2319_velmaps} indicate selected features in either OH absorption or [CII] emission more blueshifted than are detectable in spatially-integrated spectra.  The channel maps of SPT2319-55 confirm that its OH absorption extends to more blueshifted velocities than its [CII] emission in the same manner as in its spatially-integrated spectrum.  Moreover, three other sources (SPT0418-47, SPT0441-46, and SPT2132-58) exhibit more blueshifted OH absorption in channel maps than is detectable in their spatially-integrated spectra, of which two (SPT0418-47 and SPT0441-46) also exhibit more blueshifted [CII] emission in a corresponding comparison.  In one of these two sources (SPT0441-46), the OH absorption extends to more blueshifted velocities than the [CII] emission, whereas in the other (SPT0418-47), the OH absorption and [CII] emission extend to the same blueshifted velocities albeit at different positions within the Einstein ring.  Thus, in total, three sources exhibit more blueshifted OH absorption than [CII] emission in channel maps, compared with only one (of the same three) in spatially-integrated spectra, demonstrating the importance of utilising channel maps rather than relying on spatially-integrated spectra alone to identify possible outflows.
	
	To add to the evidence for separate kinematic components, we searched the channel maps for features in OH absorption that show a different spatio-kinematic pattern than [CII] emission.  This approach has the additional advantage of permitting the identification of separate kinematic components in OH absorption even in channels with detectable [CII] emission.  Before delving into this search, we first note that a one-to-one comparison between OH absorption and [CII] emission is complicated by the fact that the strength of OH absorption -- unlike that of the [CII] emission -- depends in part on the brightness of the background dust continuum, which can have a different surface brightness distribution than that of the [CII] emission.  Indeed, Figures \ref{fig:spt0202_velmaps}(a) to \ref{fig:spt2319_velmaps}(a) show such differences in some of the target sources (e.g. SPT0418-47 in Figure \ref{fig:spt0418_velmaps}(a) and SPT2132-58 in Figure \ref{fig:spt2132_velmaps}(a)), where cyan contours indicate the integrated [CII] emission and red background the dust continuum.  Thus, we should not expect the morphology of the OH absorption to perfectly match that of the [CII] emission in individual channels even if both lines trace the same gas in the disc.  In the channel maps of SPT2319-55, which shows by far the strongest and the only significant blueshifted OH absorption in spatially-integrated spectra, the OH absorption in the south of the Einstein ring indicated by the red arrows in Figures \ref{fig:spt2319_velmaps}(i) to \ref{fig:spt2319_velmaps}(l) can be seen to track the [CII] emission at velocities close to systemic where both move roughly counter-clockwise around the Einstein ring with decreasing velocity (between Figure \ref{fig:spt2319_velmaps}(l) and \ref{fig:spt2319_velmaps}(k)).  At more strongly blueshifted velocities (between Figure \ref{fig:spt2319_velmaps}(j) and \ref{fig:spt2319_velmaps}(i)), however, the OH absorption appears to reverse this behavior and move clockwise, whereas the [CII] emission does not.  This behavior suggests the presence of a separate kinematic component traced in OH absorption that moreover extends to higher blueshifted velocities than gas traced in [CII] emission contained in a rotating disc, making a strong case for a molecular outflow in SPT2319-55.
	
	In SPT0202-61, for which [CII] emission is detectable to even higher blueshifted velocities than OH absorption in both spatially-integrated spectra and channel maps, the [CII] emission at the north of the Einstein Ring as seen in the channel maps of Figure \ref{fig:spt0202_velmaps} moves counter-clockwise from panels (h) to (e).  By contrast, the patch of OH absorption indicated by the red arrows moves slightly in the opposite direction, suggesting that this feature comprises a separate kinematic component.  As this patch of OH absorption coincides with relatively weak dust continuum, it does not appear merely as a consequence of bright continuum.
	
	SPT0418-47 also shows no significant OH absorption more blueshifted than any [CII] emission in either spatially-integrated spectra or channel maps.  As can be seen in the channel maps for this source in Figure \ref{fig:spt0418_velmaps}, the [CII] emission in the northern portion of the Einstein ring moves clockwise starting around the systemic velocity in panel (f) to more strongly blueshifted velocities in panels (e) and (d).  By contrast, the peak of the OH absorption -- indicated by red arrows -- moves in the opposite direction, again indicating the presence of a separate kinematic component.  
	
	\section{DISCUSSION}
	\label{section:discussion}
	
	\subsection{Outflow inferences}
	\label{section:outflowinf}
	
	After determining the precision of corrections for atmospheric absorption as well as explicitly quantifying the measurement uncertainty in each spectral channel, we find that only one of the five sources that we analysed (SPT2319-55) shows statistically significant ($\ge 3\sigma$) evidence for OH absorption more blueshifted than [CII] emission in its spatially-integrated spectrum.  These results constrast with those reported by \citetalias{spilker18} and \citetalias{spilker20a}, who argue that four of the same five sources display OH absorption more blueshifted than [CII] emission in spatially-integrated spectra, though without explicitly quantifying the statistical significance of this absorption.  On the other hand, we find all five sources to exhibit evidence for a component in OH absorption different from that in [CII] emission in their channel maps (which were not included in the analysis of \citetalias{spilker18} and \citetalias{spilker20a}).
	
	An inspection of channel maps for SPT2319-55 reveals, in addition to OH absorption more blueshifted than [CII] emission detected in its spatially-integrated spectrum, a component in OH absorption having a different spatio-kinematic pattern than [CII] emission even in channels where both OH and [CII] are detected.  This source shows the strongest evidence for a molecular outflow.  Both SPT2132-58 and SPT0441-46 exhibit OH absorption more blueshifted than [CII] emission in channel maps that is diluted below the detection threshold in spatially-integrated spectra, thus also providing evidence for outflows.  SPT0202-61 and SPT0418-47, despite not displaying significant OH absorption more blueshifted than [CII] emission in either spatially-integrated spectra or channel maps, both exhibit features in OH absorption having different spatio-kinematic patterns than [CII] emission, and therefore evidence for separate kinematic components in OH absorption plausibly associated with outflows.
	
	\subsection{Outflow properties}
	\label{section:outflowprop}
	
	In \cite{spilker20b}, the mass outflow rate for each object thought to host an outflow is estimated using the depth of the blueshifted OH absorption.  Assuming initially the absorption to be optically thin, the OH column density is measured from the depth of the blueshifted OH absorption integrated over velocities more blueshifted than $ \SI{-200}{km.s^{-1}} $.  The total gas column density is then computed assuming an OH abundance based on the study of a star-forming region in our Galaxy, Sgr B2 \citep{gc02}.  Mass outflow rates are estimated using this column density, assuming a time-averaged thin shell geometry \citep{rupke05} with inner radius set equal to the effective radius of the dust continuum estimated using source-plane reconstructions of the gravitationally-lensed images of each source.  To allow for the possibility that the OH absorption is optically thick, an empirical correction factor to the mass outflow rates is then introduced, based on observations of 14 nearby quasars and ultra-luminous infrared galaxies (ULIRGs) in multiple OH transitions (with P Cygni profiles or high-velocity absorption wings) for which radiative transfer models have been made by \cite{ga17}.  Specifically, the correction factor is based on a correlation in this sample between mass-outflow rate and the ``maximum" blueshifted velocity of OH (defined as the velocity above which $ 98\% $ of absorption takes place).  

	In addition, \cite{spilker20b} estimate alternative mass-outflow rates based on correlations in the \cite{ga17} sample between mass-outflow rate and equivalent width of blueshifted OH absorption beyond $ \SI{-100}{km.s^{-1}} $, as well as equivalent width of blueshifted OH absorption beyond $ \SI{-200}{km.s^{-1}} $ multiplied by $ L_{IR}^{1/2} $, where $ L_{IR} $ is the total infrared luminosity corrected for gravitational lensing.  Lastly, \citet{spilker20b} apply a dimensionality-reduction technique similar to a principal component analysis to find an empirical correlation between the measured mass-ouflow rates and combinations of other parameters in the \cite{ga17} sample.  We find the same nominal depths of putative blueshifted OH absorption as those found by \citetalias{spilker18} and \citetalias{spilker20a} within measurement uncertainties.  However, having now calculated the statistical significance of these absorption features, we emphasise that the putative mass-outflow rates -- derived solely from the spatially-integrated spectra -- must have substantially higher uncertainties than presented in \cite{spilker20b}, in at least some cases rendering them consistent with zero.

	\subsection{Improving observing strategies}
	\label{section:futobs}
	
	Our analysis highlights the attention that must be paid to atmospheric absorption features when searching at sub-mm wavelengths for spectral signatures of high-$ z $ outflows traced in absorption.  Observers must not only carefully check the results of the ALMA calibration pipeline to assess the accuracy at which corrections for atmospheric absorption have been carried out, but also recognize the fact that large changes in atmospheric opacity across the bandpass lead to correspondingly large changes in the thermal noise and hence measurement uncertainties across the observed spectra.
	
	Observers who want to better verify that atmospheric absorption has been fully removed by the ALMA pipeline (improving over the test made above from the secondary calibrator) can do so by planning observations accordingly.  As observations are performed in two sidebands with one placed on the spectral line of interest and the other normally used to measure the line-free continuum, the latter sideband can be placed instead on a strong atmospheric O$ _3 $ line.  The spectral windows in this sideband will then act as a control case for the OH-containing spectral windows, and can be checked to see whether any residual features associated with the atmospheric line remain after calibration.  This check, however, may not be possible in many cases owing to restrictions on the placement of spectral windows allowed by ALMA.  In such cases, the only option left is to use the spectrum of the secondary calibrator to ensure that the degree to which corrections to atmospheric absorption have been made is likely smaller than the degree of absorption seen in the source (as conducted in our work).  The downside of this approach is that, in order for this test to be effective, the secondary calibrator spectrum must be observed at a significantly higher signal-to-noise than the target.  Therefore, a relatively large amount of observing time may need to be spent on the secondary calibrator if the closest one to the target is relatively dim.
	
	A separate issue that further complicates the detection of low-signal-to-noise outflows is the estimation of the continuum level in the target spectra.  As shown above in the case of SPT2132-58, targets with broad absorption features may have few, if any, channels to use for continuum estimation in the spectral windows covering the OH line.  In this situation, the other sideband could be used to estimate the continuum level, as is the approach taken in \citetalias{spilker18} and \citetalias{spilker20a}, but such estimates are subject to additional uncertainties introduced by the uncertain spectral index of the target source as well as errors in bandpass calibration.  In the unfortunate case of SPT2132-58, the atmospheric transmission in the continuum sideband (which has a fixed offset of \SI{12}{GHz} with respect to the OH sideband) is very low whether it is placed at either a lower or higher frequency than the target bandpass (see Figure \ref{fig:bandpasses}), so no useable continuum spectrum can be produced at all given the restrictions on the ALMA correlator set-up.  Furthermore, the OH-containing sideband cannot be extended to cover a higher frequency range since it is already at the maximum possible width.  The only remaining solution is to observe with two different correlator set-ups, with one set-up being used to measure the continuum level and the other the OH absorption.   Although doubling the observing time, there may be no other choice but to adopt this strategy in ALMA band 8 owing to deep and closely-spaced H$ _2 $O atmospheric absorption lines throughout this band (progressively less severe in lower-frequency bands).
	
	\section{SUMMARY AND CONCLUSION}
	\label{section:sum}
	
	We re-analysed observations of the five DSFGs from \citetalias{spilker18} and \citetalias{spilker20a} for which both OH $ \SI{119}{\um} $ and [CII] $ \SI{158}{\um} $ images are available at comparable angular resolutions with ALMA.  Such observations are complicated by strong absorption lines from the Earth's atmosphere imprinted onto the spectra of all targeted objects, leaving two effects that must be carefully assessed: (i) residual atmospheric absorption even after standard bandpass calibration procedures; and (ii) varying measurement uncertainties over the spectra owing to varying atmospheric opacity.  We therefore checked the precision to which corrections for atmospheric (together with instrumental) effects in the bandpass have been made, and also computed explicitly the measurement uncertainties in each velocity channel.  
	
	Extracting spatially-integrated spectra in the same manner as \citetalias{spilker18} and \citetalias{spilker20a}, we searched for OH absorption extending to more blueshifted velocities than detectable [CII] emission -- the signature used by both \citetalias{spilker18} and \citetalias{spilker20a} to identify an outflow, indicating a kinematic component different from that of the [CII] emission attributed to the galactic disc.  Over the velocity range selected to maximise the depth of any putative blueshifted OH absorption, we found the bandpass to have been corrected to within $ \lesssim 0.3 \sigma $ for all sources except SPT0418-47, for which the bandpass is corrected to $ \sim 1\sigma $.  Thus informed, we found only one source (SPT2319-55) to show statistically significant ($\ge 3\sigma$) evidence for blueshifted OH absorption in spatially-integrated spectra -- in contrast to \citetalias{spilker18} and \citetalias{spilker20a}, who found four of the five sources examined here to show unambiguous signatures of outflows, though without explicitly quantifying the statistical significance of the putative blueshifted OH absorption.
	
	In spatially-integrated spectra, signals that do not span the entire source can be diluted to below detectability, whether it be in OH absorption or [CII] emission.  To guard against this possibility, we made channel maps in both OH absorption and [CII] emission to find signals that are even more blueshifted than are detected in spatially-integrated spectra.  Furthermore, to find additional evidence for a separate kinematic component that can possibly arise from an outflow, we searched for an OH component exhibiting a different spatio-kinematic pattern than the [CII] emission.  The full results for each source are summarised below.
	\\
	
	\noindent
	SPT0202-61:  no OH absorption blueshifted beyond the [CII] emission in its spatially-integrated spectrum (in agreement with \citetalias{spilker20a}) or in channel maps; on the other hand, component in OH absorption having different spatio-kinematic pattern than [CII] emission in channel maps.
	\\ \\
	\noindent
	SPT0418-47:  tentative ($2.4\sigma$) OH absorption blueshifted beyond detactable [CII] emission in spatially-integrated spectrum; over the velocity range of the tentative blueshifted OH absorption, upper limit on precision of bandpass correction corresponding to $\sim 1 \sigma $ of the measurement uncertainty, and may therefore contribute in part to this tentative absorption feature; no OH absorption blueshifted beyond the [CII] emission in channel maps, but a clear component in OH absorption having different spatio-kinematic pattern than [CII] emission.
	\\ \\
	\noindent
	SPT0441-46:  tentative ($ 2.7\sigma $) blueshifted OH absorption in its spatially-integrated spectrum; significant OH absorption blueshifted beyond any [CII] emission in channel maps.
	\\ \\
	\noindent
	SPT2132-58:  tentative ($ 2.2\sigma $) blueshifted OH absorption in its spatially-integrated spectrum; significant OH absorption blueshifted beyond any [CII] emission in channel maps.
	\\ \\
	\noindent
	SPT2319-55:  highly significant ($6.1 \sigma$) blueshifted OH absorption its spatially-integrated spectrum, also seen in channel maps; possible component in OH absorption having different spatio-kinematic pattern than [CII] emission in channel maps.
	\\
	
	\noindent In brief, we find evidence of separate kinematic components plausibly associated with molecular outflows in all five of the sources included in our study.
	
	These results highlight the potential pitfalls in relying on blueshifted OH absorption in spatially-integrated spectra alone to infer the presence of candidate outflows, and the importance of utilising the full spatio-kinematic information provided by the channel maps to search for and verify candidate outflows.  Deeper follow-up observations are clearly warranted for the four sources that we find to exhibit evidence for molecular outflows, which if confirmed also permit a more precise determination of their outflow properties.
	
	Future studies should make sure -- as we have -- that atmospheric absorption is fully corrected for at the level of any putative outflow absorption, and that the varying noise level across spectral channels is properly computed.  In addition, the continuum level should be better measured where possible in those sources where the putative OH absorption spans most of the bandpass.  As can be seen in Figure \ref{fig:bandpasses}, two of the remaining four sources found to show signs of outflows presented in \citetalias{spilker20a} based on OH absorption more blueshifted than detetcable CO emission (SPT0544-40 and SPT2311-54) also contain large changes in atmospheric absorption across the OH-containing bandpasses, requiring a proper accounting of the measurement uncertainties at the velocities of the putative blueshifted OH absorption.  We recommend that, for observations in spectral lines, example scripts for data calibration and analysis of ALMA data in CASA include descriptions for assessing the accuracy of bandpass calibration and computing the measurement uncertainties in individual spectral channels.
	
	\section{Acknowledgments}
	
	J. Nianias and J. Lim acknowledge support from the Research Grants Council of Hong Kong for conducting this work under the General Research Fund 17312122. M. Yeung acknowledges support from the Research Grants Council of Hong Kong for conducting this work under the General Research Fund 1730451.  This paper makes use of the following ALMA data: ADS/JAO.ALMA\#2018.1.00191.S, ADS/JAO.ALMA\#2016.1.01499.S. ALMA is a partnership of ESO (representing its member states), NSF (USA) and NINS (Japan), together with NRC (Canada), MOST and ASIAA (Taiwan), and KASI (Republic of Korea), in cooperation with the Republic of Chile. The Joint ALMA Observatory is operated by ESO, AUI/NRAO and NAOJ.  We acknowledge the assistance provided by the ALMA helpdesk East Asia.  We also thank the anonymous referee for their constructive feedback.
	
	\software{CASA \citep{casaref}, Astropy \citep{astropy:2013, astropy:2018, astropy:2022}, Matplotlib \citep{mplref}, ESSENCE \citep{tsukui23}}
	
	\bibliography{mybib}
	
\end{document}